\documentclass[a4paper]{article}

\title{Correlation of internal representations in feed-forward neural networks}

\author{A. Engel\thanks{email: andreas.engel@physik.uni-magdeburg.de}\\[1ex]
        \small Institut f\"ur Theoretische Physik\\
				\small Otto-von-Guericke Universit\"at, Postfach 4120, D-39016 Magdeburg, Germany
			 }

\usepackage{amstex,epsfig,graphicx}
\newcommand{\ksi}{\boldsymbol{\xi}^{\mu}_k}
\newcommand{\J}{\mathbf{J}_k}
\newcommand{\sgn}{\text{sgn}}
\newcommand{\Tr}{\text{Tr}}
\newcommand{\KK}{\frac{K-1}{2}}

\begin{document}

\maketitle

\begin{abstract}
Feed-forward multilayer neural networks implementing random input-output
mappings develop characteristic correlations between the activity
of their hidden nodes which are important for the
understanding of the storage and generalization performance of the network.
It is shown how these correlations can be calculated from the joint
probability distribution of the aligning fields at the hidden units for
arbitrary decoder function between hidden layer and output. Explicit results
are given for the parity-, and-, and committee-machines with arbitrary
number of hidden nodes near saturation.
\end{abstract}

\vspace{0.5cm}

Multilayer neural networks (MLN) are powerful information processing
devices. Because of their computational abilities they are 
the workhorses in practical applications of neural networks and a lot of
effort is devoted to a thorough understanding of their functional principles. 
At the same 
time their theoretical analysis within the framework of statistical mechanics 
is much harder than that for the single-layer perceptron.
It was realized from the beginning that the properties of the internal
representations  defined as the activity patterns of the hidden units resulting
from certain inputs  are crucial for the understanding of the storage and
generalization abilities of MLN \cite{MePa, GriGro, Priel, Scho, MoZe}.
Qualitatively the flexibility of MLN stems from the fact that the different
subperceptrons between input and hidden layer can share the effort to
produce the correct output. This {\em division of labour} gives rise to
particular correlations between the activity of the hidden nodes. Near
saturation these correlations become a characteristic feature  of the 
decoder function between hidden units and output of the MLN under
consideration and determine different aspects of its performance.

Several ad-hoc approximations have been used to calculate these correlations, 
e.g., it was assumed that all internal
representations giving the correct output (so called legal internal
representations, LIR) are equiprobable \cite{GriGro,Priel} or that only 
internal representations at the decision boundary of
the decoder function occur \cite{Priel}. In the present letter we show how these
correlations between the hidden units can be
calculated for a MLN of tree-architecture and give explicit results for the
parity- (PAR), and- (AND) and committee- (COM) machine with arbitrary 
number $K$ of hidden nodes near saturation. 

A MLN of tree-architecture is given by $N$ input nodes $\xi_{ik}$ grouped into $K$ sets of
$N/K$ nodes each, $K$ hidden nodes $\tau_k$ and one output $\sigma$. The
inputs $\boldsymbol{\xi}_k=\{ \xi_{ik}, i = 1, \ldots, N/K\}$ are coupled to
the $k$-th hidden unit by spherical couplings
$\J=\{ J_{ik} \in \Bbb{R}, i = 1, \ldots, N/K,\; \J^2
= N/K\}$
according to $\tau_k = \sgn (\J \boldsymbol{\xi}_k)$. 
Each hidden node has therefore
its own set of inputs (non-overlapping receptive fields). The hidden units
$\tau_{k}$ determine the output through a fixed Boolean function 
$F(\{\tau_{k}\})$. A set of input--output mappings $\{\ksi,\sigma^{\mu}\}, \mu = 1, \ldots, p$ 
is generated at random where each bit is $\pm 1$ with equal probability. 
The couplings $\J$ are then adjusted in such a way that the MLN gives the
desired output $\sigma^{\mu}$ for each input $\ksi$. This is generically possible 
only if $p/N =\alpha  < \alpha_{c} $. 

We are interested in the correlations
\begin{equation}
c_n = \frac{1}{\alpha N} \sum^{\alpha N}_{\mu = 1} \tau^{\mu}_{1}
\tau^{\mu}_2 \tau^{\mu}_3 \ldots \tau^{\mu}_{n} 
\end{equation}
near saturation, i.e. for $\alpha \to \alpha_c$. From the statistical
properties of the inputs it follows that
\begin{equation}
c _{n} = \langle\langle \tau_{k_1} \tau_{k_2} \ldots \tau_{k_n} \rangle\rangle
\end{equation}
where $\langle\langle \ldots \rangle\rangle$ denotes the average over the 
input-output pairs and ${k_1,\ldots, k_n}$ is any set containing $n$ different 
natural numbers between $1$ and $K$.

The $c_n$ can be calculated from the joint probability distribution of
internal representations
\begin{equation}
P(\tau_1, \ldots, \tau_k) = \langle\langle \frac{\int \prod^K_{k = 1}
d\mu(\J)\prod^K_{k = 1} \theta (\tau_k \J \boldsymbol{\xi}_k^1) \prod_{\mu} \theta 
(\sigma^{\mu} F(\{\sgn (\J \ksi\}))}
{\int \prod^K_{k = 1} d\mu(\J)\prod^K_{k = 1} \prod_{\mu} \theta 
(\sigma^{\mu} F(\{\sgn (\J \ksi)\}))}\rangle\rangle
\end{equation}
The calculation of $P(\tau_1, \ldots, \tau_k)$ parallels the determination of the 
local aligning field distribution for the perceptron \cite{KeAb,Ga89} (see also \cite{BHS,
GriGro, Gri}). The general result within replica symmetry is
\begin{equation}\label{RS}
P (\tau_1, \ldots, \tau_k) = \langle\langle \delta_{\sigma, F (\{\tau_k\})} 
\int \prod_k Dt_k
\frac {\prod_k H (Q t_k \tau_k)}{ \Tr_{\eta_k}' \prod_k H(Q t_k \eta_k)}
\rangle\rangle_{\sigma}
\end{equation}
where $\delta_{n,m}$ is the Kronecker symbol and the primed trace 
$\Tr_{\eta_k}' = \Tr_{\eta_k} \delta_{\sigma, F(\{\eta_k\})}$
is restricted to the legal internal representations.
Moreover $Dt = e^{-t^2/2}dt/\sqrt{2 \pi}$ and $H (x)=\int^{\infty}_{x} Dt$ 
as usual. We do not display the saddle-point equation
necessary to determine $Q = \sqrt{q/(1-q)}$ as a function of $\alpha$
since we are mainly interested in the saturation limit $\alpha \to \alpha_c$
implying $q \to 1$ and therefore $Q \to \infty$. 

In this limit the integrand in (\ref{RS}) either tends to zero or to one depending
on the values of the $t_k$.
When calculating $P (\tau_1, \ldots, \tau_k)$ explicitely for small $K$
and special decoder functions one realizes a simple general rule. Consider
the system before learning. All internal representations have an a-priori 
probability $2^{-K}$. Those already compatible with the desired output are
not modified, all the others are shifted by the learning process {\em to the
nearest decision boundary} of the decoder function. This is reminiscent of
the aligning field distribution of the simple perceptron \cite{Ga89, Op88}
and has a natural interpretation within the cavity approach \cite{Gri,
Wong}. On the basis of this general rule it is possible to
determine $P(\tau_1, \ldots, \tau_k)$ for arbitrary $K$ and arbitrary
decoder function.

As examples we derive in the following explicit results for the PAR-,
AND- and COM-machines defined by the decoder functions 
$F (\{\tau_k\}) =\prod_k \tau_{k}, F(\{\tau_k\}) = \sgn (\sum_k \tau_k - K + 1/2)$ and
$F(\{\tau_k\}) = \sgn (\sum_k \tau_k)$ respectively. For the PAR- and COM-machine 
we can set all outputs equal to $+1$ without loss of generality for 
symmetry reasons whereas for the AND-machine we
have to stick to random outputs $\sigma^{\mu} = \pm 1$ with equal
probability. 

In the case of the PAR-machine all internal representations are at the
decision boundary of the decoder function. Hence all LIR gain in addition
to their a-priori weight $2^{-K}$ an equal share from the $2^{1-K}$ internal
representations that are eliminated by the learning process. Therefore for
$\alpha \to \alpha_c$ all LIR have equal probability $2^{1-K}$ which results
in $c_n = 0$ for all $n = 1, \ldots, K-1$ and $c_K = 1$.

In the case of the AND-machine there is only one LIR for the output $\sigma=+1$,
namely $\tau_1 =\ldots= \tau_K=1$. It contributes $1/2$ to all $c_n$. If
$\sigma = - 1$ all but one internal representations are LIR. Only those
with exactly one $\tau_k = - 1$ are at the decision boundary of the decoder
function and consequently only their probability is changed by the learning
process. For symmetry reason it is clear that all of them get on equal share
$2 ^{-K}/K$ from the elimination of the internal representation
$\tau_1 = \ldots = \tau_K = + 1$ in addition to their a-priori weight 
$2^{-K}$. The calculation of the resulting contribution from $\sigma = -1$ to the
correlations $c_n$ can be most easily accomplished by observing that $c_n =0$ 
for all $n$ before learning. To calculate $c_n$ after learning one
has hence only to take into account those LIR with exactly one $\tau_k = -1$. The
result is $c_n = \frac{1}{2} - n 2^{-K}/K$. As expected all
correlations are dominated by the restrictive case $\sigma = +1$ of the output.

For the COM-machine the calculation is more involved. As usual we only
consider odd  values of $K$. The decision boundary is given by all LIR with
$\sum _k \tau_k = 1$. All these gain an equal share from the $2^{1-K}$
internal representation that have to be eliminated by the learning. Hence
\begin{xalignat*}{2}
P(\{\tau_k\}) &= 2^{-K}&\text{if}\qquad &\sum_k \tau_k > 1\\
P (\{\tau_k\})&= 2^{-K} + \frac{1}{2}\left[\binom{K}{\KK}\right]^{-1}&
\text{if}\qquad &\sum_k \tau_k= 1\\
P (\{\tau_k\}) &= 0   &\text{else} &  
\end{xalignat*}
To determine the values of $c_n$ from this $P(\{\tau_k\})$ it is convenient
to consider the contribution from the regular part 
$P^{(r)}(\{\tau_k\}) =2^{-K}$ if $\sum_k \tau_k \geq 1$ and that from the extra part 
$P^{(e)}(\{\tau_k\}) = \frac{1}{2} [\binom{K}{\KK}]^{-1}$ for 
$\sum_k \tau_k = 1$ seperately. 
The regular part contribution to even moments is zero due to symmetry. 
Its contribution to odd moments is
\begin{equation}\label{reg}
c_n^{(r)} = 2^{-K} \sum^{\KK}_{m = 0} \sum^n_{i = 0} (-1)^i
          \binom{n}{i}\binom{K-n}{m-i}
\end{equation}
since $i = 0, \ldots, n$ of the $m = 0, \ldots, \KK$ minus ones of a 
LIR can be found in $\tau_1, \ldots, \tau_n$ whereas the remaining $(m-i)$
minus ones are to be distributed between the remaining $(K-n)$ 
$\tau_{n+1}, \ldots,\tau_k$. After some algebra using 
properties of binomial coefficents \cite{Knuth} this can be simplified to
\begin{align}
c_n^{(r)}& = 2^{-K} \sum^{n-1}_{i = 0} (-1)^{i}
\binom{n-1}{i}\binom{K-n}{\KK-i}\\
    &= 2^{-K} (-1)^{\frac{n-1}{2}} \frac{(n-2)!!}{(K-2)(K-4) \ldots (K-n+1)}
           \binom{K-1}{\KK}\\
  &= 2^{-K} \frac {\Gamma(\frac{n}{2})\; \Gamma(1- \frac{K}{2})\; \Gamma (K)}
  {\sqrt{\pi}\; \Gamma (\frac{n-K+1}{2})\;[ \Gamma(\frac{K+1}{2})]^2}
\end{align}
Similarily one gets for the extra part contribution
\begin{equation}\label{extra}
c_n^{(e)} = \frac{1}{2} \left[ \binom{K}{\KK}\right]^{-1}
            \sum^n_{i = 0} (-1)^i \binom{n}{i} \binom{K-n}{\KK-i}
\end{equation}
which results in
\begin{align}
c_n^{(e)} &= \frac{1}{2} (-1)^{\frac{n-1}{2}}\frac{n!!}{K(K-2)(K-4)
  \ldots (K-n+1)}\\
   &= \frac{\Gamma(\frac{n}{2} + 1)\; \Gamma (1 - \frac{K}{2})}
	 {2 \sqrt{\pi}\;\Gamma(\frac{n - K + 1}{2})}
\end{align}
for $n$ odd and
\begin{equation}
c_{n}^{(e)} = - c_{n-1}^{(e)}
\end{equation}
if $n$ is even.
The final result for the correlations of the COM -machine is hence
\begin{xalignat}{2}
c_n(K) &= \frac{\Gamma(\frac{n + 1}{2})\; \Gamma (- \frac{K}{2})}
    {2\; \sqrt{\pi}\; \Gamma(\frac{n-K}{2})} &\text{if } & n\; \text{even}\label{even}\\
c_n(K) &= \frac{\Gamma (\frac{n}{2})\; \Gamma (1-\frac{K}{2})}
    {\sqrt{\pi}\; \Gamma(\frac{n-K+1}{2})}\left[\frac{\Gamma (K)}
		{2^K\; [\Gamma(\frac{K+1}{2})]^2} + \frac{n}{2 K}\right] &\text{if } &
                n\;\text{odd}\label{odd}
\end{xalignat}
Note that for $n$ even one has $c_{K-n+1}=(-1)^{(K+1)/2} c_n$. As an example
these results are shown in the figure for $K = 25$.
\epsfig{bbllx=1,bblly=350,bburx=700,bbury=500,file=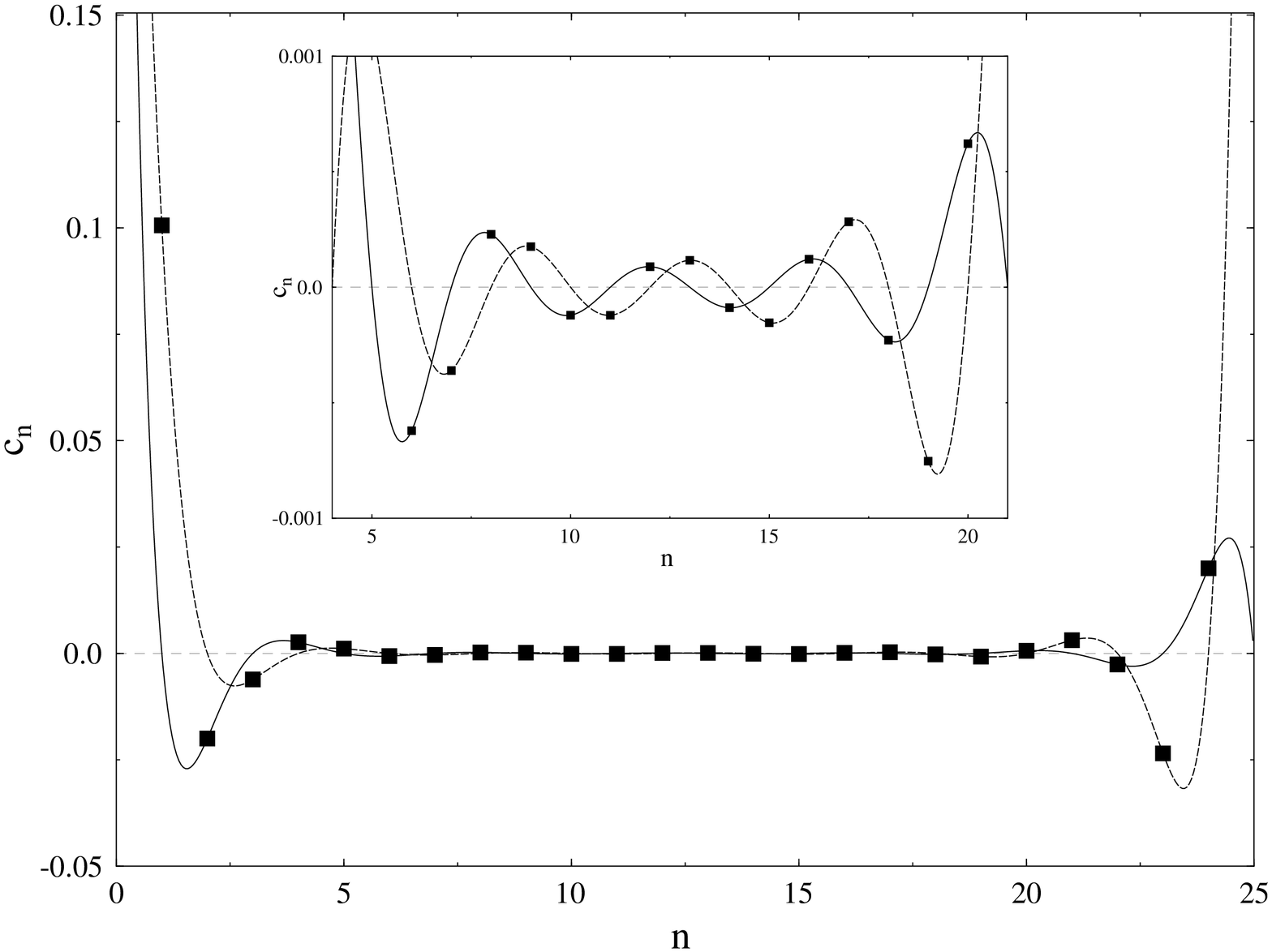,width=10cm,angle=-90}
\vspace*{-1cm}

\begin{minipage}[b]{11cm}{\footnotesize Moments of internal representations for a
               committee machine with $K=25$ hidden units. The symbols are
               the results for integer $n$ following from eqs.(\ref{reg}) and
               (\ref{extra}), the full and dotted line are given by eqs.(\ref{even})
               and (\ref{odd}) respectively. The inset shows an enlarged region
							 of the plot.}
\end{minipage}

It is straightforward to obtain the asymptotic behaviour of the moments for
$K\to\infty$. For the COM-machine moments $c_n$ with either $n$ or $K-n$
small remain the largest ones in this limit. Explicitly one gets with the
abbreviation $C=1/\sqrt{2\pi K}$
$c_1\approx C, c_2=-1/(2K)$, $ c_3\approx -C/K, c_4=3/(2K(K-2)), c_5\approx 3C/K^2$ and
$c_K\approx (-1)^{(K-1)/2}$, $ c_{K-1}=(-1)^{(K-1)/2}/(2K), c_{K-2}\approx
(-1)^{(K+1)/2}/(2K), c_{K-3}=(-1)^{(K+1)/2} 3/(2K(K-2)), c_{K-4}\approx
(-1)^{(K-1)/2} 3/(2K^2)$. 

So far we have considered a MLN with fixed decoder function and have
determined the correlations $c_n$ resulting near saturation.
It is tempting to investigate also the complementary question and 
to determine the storage capacity of an ensemble of $K$ uncoupled
perceptrons with prescribed correlations $c_n$. For the COM-machine it
is, e.g., known that already from the prescription of $c_1$ alone one gets
the correct RS-asymptotics $\alpha_c\cong K^2$ for the storage capacity
(which is, however, known to be unstable with respect to RSB). It is
interesting to see whether the inclusion of other correlations can alter
this asymptotics \cite{MEK}.

Finally it should be noted that the results obtained in this letter rely
on the assumption of replica symmetry
for  the determination  of the aligning field distribution whereas it is well
known that replica symmetry breaking is crucial for the calculation of the
storage capacity of MLN. On the other hand it is merely the {\em qualitative}
behaviour of the aligning field distribution that is important for the
determination of the $c_n$. Since this is
known to be hardly modified by RSB \cite{ET,MEZ} it seems likely that the
results for the correlations will not be significantly altered by the
inclusion of RSB. 

\bibliographystyle{unsrt}

\end{document}